\pdfoutput=1
\documentclass[11pt,a4paper]{article}

\usepackage[english]{babel}
\usepackage[T1]{fontenc}
\usepackage[utf8]{inputenc}

\usepackage{amsmath,amssymb,amsfonts,mathtools}
\usepackage{geometry}
\usepackage{graphicx}
\usepackage{float}
\usepackage{microtype}
\usepackage{caption}
\usepackage{hyperref}

\usepackage{mathptmx}

\graphicspath{{./}}
\makeatletter
\renewcommand{\maketitle}{
\begin{flushleft}
{\LARGE \bfseries \@title \par}
\vspace{1em}
{\normalsize \@author \par}
\vspace{1em}
\end{flushleft}
}
\makeatother

\title{
From Organization to Viability:\\
A Multi-Level Analysis of Gait Dynamics Under Occlusal Constraint
}

\author{
Jacques Raynal$^{1,*}$,
Pierre Slangen$^{2}$,
Elsa Raynal$^{3}$,
Jacques Margerit$^{4}$\\[0.5em]
{\small $^{1}$Laboratory of Bioengineering and Nanosciences (LBN), University of Montpellier, France}\\
{\small $^{2}$EuroMov Digital Health in Motion, University of Montpellier, IMT Mines Al\`es, Al\`es, France}\\
{\small $^{3}$Certified Sophrologist and Dental Assistant, Sensorimotor Practice, Montpellier, France}\\
{\small $^{4}$Emeritus Professor, University of Montpellier, France}\\[0.5em]
{\small $^{*}$Corresponding author: \texttt{raynal.cab@gmail.com}}
}

\date{}

\begin{document}

\maketitle

\begin{abstract}

    Clinical interpretation often assumes that observable performance provides sufficient information about an adaptive system. However, this assumption relies on a correspondence between output metrics and system organization that may not hold. The preceding Level~3 study showed that neither an aggregated scalar score nor a static exploratory UMAP embedding uniquely resolved the occlusal observational probes. The multivariate representation retained information not expressed by scalar aggregation, but the six probes remained substantially overlapping and did not form independently validated condition-specific clusters.

    The present study extends this framework by introducing a fourth analytical level centered on observed longitudinal viability. The unresolved static non-identifiability identified at Level~3 motivates a change in analytical question: when configurations cannot be uniquely distinguished at a given session, do they nevertheless exhibit different longitudinal displacements between measurement sessions?
    
    Using an exploratory single-case design in a participant with Parkinson's disease, gait data were recorded with instrumented insoles under three occlusal observational probes: neutral natural occlusion (ONL), a nominal 2.5-degree increase in vertical dimension of occlusion (OC2.5), and a nominal 3-degree increase in vertical dimension of occlusion (OC3). Two measurement sessions were conducted eleven weeks apart, during which the participant underwent a structured sensorimotor intervention. The occlusal conditions are interpreted as discrete observational probes applied during measurement, rather than as continuous causal drivers of longitudinal evolution.
    
    A common PCA representation was used to describe M1--M2 centroid displacement. In the selected PC1--PC2 projection, OC3 exhibited the smallest Euclidean centroid displacement, ONL occupied an intermediate position, and OC2.5 showed the largest displacement. This hierarchy provides an exploratory operationalization of Level~4, in which lower longitudinal displacement is treated as a representation-dependent proxy for lower reorganization over time.
    
    These differences should be interpreted as within-subject exploratory tendencies rather than validated clinical separations. They do not establish a causal occlusal effect, a validated viability threshold, a therapeutic optimum, or a covariance-independent ranking. Level~4 provides a retrospective observational proxy for longitudinal behavior; its predictive extension is addressed separately at Level~5.
\end{abstract}


\section{Introduction}

Observable performance does not uniquely determine the organization of an adaptive system. Clinical reasoning has historically relied on measurable variables, assuming that functional outputs provide sufficient access to the internal structure of the system. In gait and postural assessment, parameters such as walking speed, cadence, stride length, asymmetry, and center-of-pressure behavior are commonly used to characterize performance \cite{winter1995balance,horak2006postural,delDin2016freeliving}.

However, complex biological systems may generate similar observable outputs from different internal organizations. In gait dynamics, variability is not merely noise; it often reflects the structure of motor control and adaptive regulation \cite{hausdorff2009gait,latash2012motor}.

A previous multi-level framework showed that neither an aggregated scalar score nor a static exploratory embedding uniquely resolved the occlusal observational probes \cite{raynal2026level3}. The Level~1 ranking was sensitive to score construction, while the Level~3 UMAP representation showed substantial overlap rather than independently separated condition-specific clusters. This persistent ambiguity was described as representational non-identifiability.

A key limitation of static approaches is that neither aggregated performance nor a model-dependent low-dimensional embedding determines how the observed representation will change over time under constraint.

In the preceding framework, Level~4 was introduced only as a conceptual extension describing possible relationships among system configurations. The present work differs by proposing an operational, retrospective approximation of Level~4 based on observed displacement between two measurement sessions. This operationalization remains exploratory and does not convert representational displacement into a validated biomarker of viability.

Level~4 emerges from the unresolved insufficiency of static representations. Level~1 did not provide a weighting-independent ranking, and Level~3 did not organize the observations into clearly separated condition-specific regions. The relevant question therefore shifts from whether the configurations can be uniquely distinguished at one time point to whether their representations change differently over time.

The present study asks whether observational probes that remain overlapping or ambiguous in a static representation exhibit different M1--M2 centroid displacements within a common PCA space. This question cannot be answered by Level~1 metrics alone or by a Level~3 embedding considered at one session. It requires a fourth analytical level centered on longitudinal behavior.

The predictive extension of this framework is developed separately as Level 5, where observed longitudinal viability becomes the target of an internal predictive approximation in selected PCA representation \cite{raynal2026level5}. In this exploratory study, the concept is examined using gait data from a Parkinsonian patient measured under selected occlusal constraints before and after a structured sensorimotor intervention. The vertical dimension of occlusion (VDO) is considered as an experimentally varied constraint applied to an adaptive neuromechanical system.

The objective is not to establish causal therapeutic effects, but to formalize and illustrate how longitudinal displacement may provide complementary descriptive information when configurations remain insufficiently resolved by lower-level static representations.

This limitation can be reformulated geometrically. Two observational probes may remain substantially overlapping when evaluated through aggregated performance or a static multivariate embedding. Such overlap does not determine whether their mean representations undergo comparable changes between sessions.

The critical distinction is therefore not only where observations are located within a selected representation at one time point, but how their condition-level summaries change between M1 and M2. Overlapping configurations may exhibit different longitudinal centroid displacements without constituting discrete physiological states.

\begin{figure}[H]
\centering
\caption{\textbf{Conceptual transition from Level~3 to Level~4.}
Level~3 provides a static exploratory multivariate representation but does not necessarily identify separated condition-specific states. Observational probes may remain substantially overlapping at a given session. Level~4 introduces a longitudinal criterion by comparing how condition-level representations change between M1 and M2 within a common projection. In the present framework, lower centroid displacement is used as a retrospective, representation-dependent proxy for lower longitudinal reorganization. The diagram is conceptual and does not represent a validated physiological state space.}
\label{fig:concept}
\includegraphics[width=0.95\textwidth]{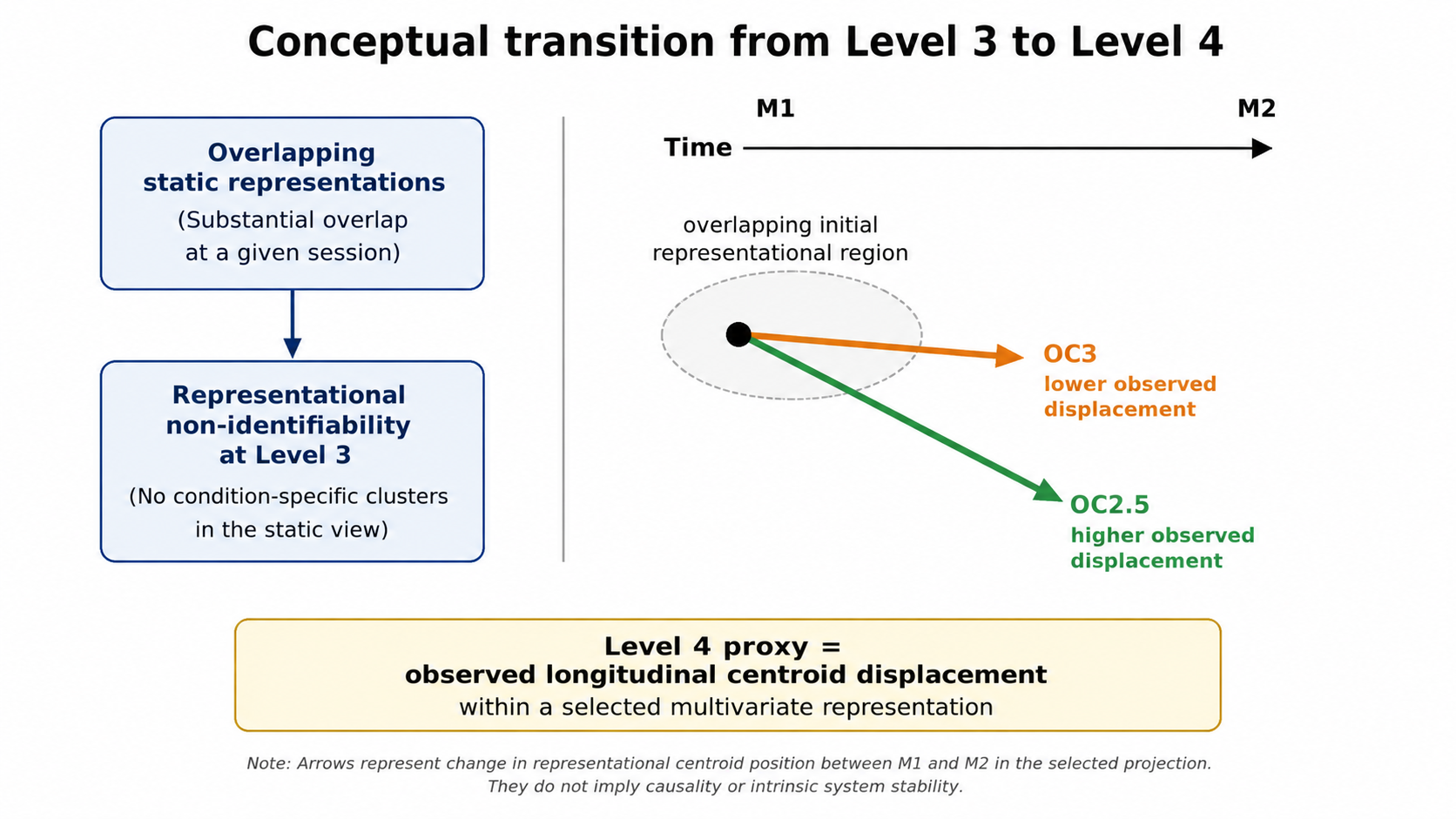}
\end{figure}

Figure~\ref{fig:concept} illustrates the transition from static representational non-identifiability at Level~3 to a longitudinal comparison of representational displacement at Level~4.

\section{Theoretical Framework}

\subsection{Representational non-identifiability}

Let \(x \in \mathbb{R}^{n}\) denote an observable vector of clinical and biomechanical variables, and let

\[
r = \Phi(x), \qquad r \in \mathbb{R}^{m},
\]

denote a model-dependent representation obtained through aggregation or dimensionality reduction.

Representational non-identifiability arises when different observational probes cannot be uniquely resolved by the selected representation. This may occur when their aggregated scores are similar, when their multivariate distributions overlap, or when their relative ordering depends on preprocessing, weighting, or projection choices.

Formally, for two observational probes \(\lambda_1\) and \(\lambda_2\), non-identifiability may be expressed as

\[
P(r \mid \lambda_1)
\approx
P(r \mid \lambda_2),
\]

or more generally when the overlap between their representational distributions is large relative to their separation.

This formulation does not assume that the representation provides direct access to an underlying physiological state. It states only that the selected analytical representation does not establish a one-to-one correspondence between the observational probe and a uniquely identified system configuration.

\subsection{From static representation to longitudinal viability}

The earlier framework distinguished three analytical levels: aggregated observable performance, conceptual dynamical organization, and exploratory multivariate embedding \cite{raynal2026level3}. The revised Level~3 analysis did not identify separated condition-specific states. Instead, it showed that static representational non-identifiability persisted despite the use of a richer multivariate representation.

Level~4 therefore does not begin from a set of empirically validated latent states. It begins from the observation that static representations remain insufficient for describing longitudinal behavior.

Let \(\lambda \in \Lambda\) denote an observational constraint and let

\[
r_{t,\lambda} = \Phi(x_{t,\lambda})
\]

denote the representation of observations recorded at session \(t\) under probe \(\lambda\).

The longitudinal change associated with probe \(\lambda\) is described by

\[
\Delta r_{\lambda}
=
\bar{r}_{M2,\lambda}
-
\bar{r}_{M1,\lambda},
\]

where \(\bar{r}_{M1,\lambda}\) and \(\bar{r}_{M2,\lambda}\) are the condition centroids within a common representational space.

The corresponding displacement is

\[
d_{\lambda}
=
\left\lVert
\bar{r}_{M2,\lambda}
-
\bar{r}_{M1,\lambda}
\right\rVert.
\]

In the present study, lower displacement is used as a retrospective and representation-dependent proxy for lower longitudinal reorganization. It is not interpreted as a direct measurement of physiological viability, stability, or therapeutic suitability.

Thus, Level~4 shifts the analysis from static overlap to longitudinal displacement. Its object is not a latent state considered in isolation, but the observed change of a condition-level representation between two sessions.

\subsection{Geometric interpretation in a common PCA representation}

The multivariate observations are projected into a common lower-dimensional space using

\[
r = \Phi_{\mathrm{PCA}}(x),
\qquad
\Phi_{\mathrm{PCA}}:
\mathbb{R}^{n}
\rightarrow
\mathbb{R}^{m}.
\]

In the present study, \(m=2\), corresponding to the PC1--PC2 plane. PCA was selected because it provides a reproducible linear transformation in which positions, centroids, and Euclidean displacements can be computed consistently across M1 and M2.

The PCA representation should not be interpreted as an intrinsic physiological manifold or as direct access to the internal state of the system. It is a model-dependent coordinate system summarizing part of the variance contained in the multivariate observations.

For each observational probe \(\lambda\), longitudinal displacement is defined as

\[
d_{\lambda}
=
\left\lVert
\bar{r}_{M2,\lambda}
-
\bar{r}_{M1,\lambda}
\right\rVert_{2}.
\]

This quantity is an ordinary Euclidean distance in the selected PCA plane. It is not a geodesic distance, does not estimate a physiological trajectory, and does not establish that the two centroids correspond to discrete biological states.

Level~4 therefore uses geometric language in a restricted descriptive sense. The analysis compares condition-level displacement within a common representation and examines whether the observed ordering is robust to resampling and covariance normalization.

A lower value of \(d_{\lambda}\) is interpreted only as lower observed centroid displacement within this representation. It is not sufficient, by itself, to establish clinical viability, physiological stability, or therapeutic superiority.

\subsection{Level~4 and the future possibility of Level~5}

The present work does not implement a prospective predictive model. It focuses on the retrospective observation of M1--M2 displacement and introduces Level~4 as an exploratory framework for describing longitudinal change within a common representation.

A subsequent Level~5 study examines whether this observed representation-dependent transformation can be internally approximated by a learned model \cite{raynal2026level5}:

\[
\widehat{r}_{t+1}
=
G_{\theta}(r_t,\lambda_t).
\]

This formulation does not imply that clinical viability can be predicted. It defines a restricted computational question: whether the displacement observed retrospectively at Level~4 can be approximated within the same single-subject dataset.

The progression is therefore from static representational non-identifiability, to retrospective longitudinal displacement, and then to internal approximation of that observed displacement.

\section{Materials and Methods}

\subsection{Study design}

The data analyzed in this work come from an exploratory intra-subject design. A Parkinsonian participant was assessed during two gait analysis sessions separated by eleven weeks. The first session is denoted M1 and the second M2. Between M1 and M2, the participant underwent eleven structured sensorimotor intervention sessions.

This design is consistent with an \(n\)-of-1 exploratory framework, allowing a detailed analysis of within-subject transformations while limiting population-level generalization \cite{smith2012singlecase,kazdin2011singlecase}. The purpose was not to establish a clinical protocol or causal treatment effect, but to examine how condition-level gait representations changed longitudinally when the participant was observed under selected occlusal probes.

\subsection{Participant and clinical context}

The participant was a Parkinsonian walker with autonomous gait. Parkinson's disease is characterized by alterations of gait, postural control, and sensorimotor integration \cite{morris2001parkinson,mirelman2019gait,delDin2016freeliving}. The neurological condition provides a clinically relevant context for studying adaptive gait dynamics, but it is not used here as an explanatory variable.

\subsection{Ethical Considerations}

The participant provided written informed consent for the use of anonymized gait data for research and publication. The study was conducted in accordance with the principles of the Declaration of Helsinki. No personally identifiable information is reported.

\subsection{Sophrology-oriented sensorimotor intervention protocol}

Between M1 and M2, the participant underwent eleven individual sessions of sophrology-oriented sensorimotor intervention. The intervention was delivered by a co-author of the present study, who is a certified sophrologist and also a dental assistant. Her role combined sophrology-based sensorimotor practice with professional familiarity with dental care and the occlusal clinical context.

The dual professional background of the practitioner is relevant to the present framework: the intervention targeted global sensorimotor regulation, while familiarity with dental care provided clinical understanding of the occlusal constraints under investigation.

Each session was structured around three main components: body awareness and interoceptive perception, postural regulation, and sensorimotor integration. The intervention combined guided attention to bodily sensations, controlled breathing, low-amplitude movement tasks, postural alignment work, and proprioceptive engagement.

The protocol was not designed to optimize a single gait metric. Rather, it aimed to modulate the participant's global sensorimotor organization. From the standpoint of the present study, this intervention constitutes a structured transformation of the system between M1 and M2. Its relevance lies in providing part of the longitudinal context between M1 and M2, within which condition-dependent representational displacement could be observed. The intervention was not isolated as a causal factor.

Mind-body and sophrology-related interventions have been discussed as approaches capable of modulating body awareness, stress regulation, and sensorimotor control, although evidence remains heterogeneous and context-dependent \cite{vanrangelrooij2020sophrology,mehling2011bodyawareness,schmalzl2014movement}.

\subsection{Occlusal conditions and rationale for selection}

The initial experimental protocol included six occlusal conditions. The present analysis focuses on three: ONL, OC2.5, and OC3.

ONL corresponds to neutral natural occlusion and serves as the baseline reference. OC2.5 and OC3 correspond to two controlled increases in vertical dimension of occlusion (VDO), respectively 2.5 degrees and 3 degrees, performed in centric relation. These two conditions were selected because they represent clinically relevant candidate configurations for exploring possible improvement in the patient.

The definition and clinical interpretation of the occlusal conditions were performed by the corresponding author, based on dental clinical expertise in occlusion.

Other conditions from the initial protocol, including clenching, open-mouth disengagement, and mandibular protrusion, were excluded from the present analysis. This restriction was intentional. Those conditions introduce additional confounding factors such as voluntary muscular contraction, occlusal disengagement, or displacement of the mandibular hinge axis. By focusing on ONL, OC2.5, and OC3, the analysis isolates a narrower question: how do two close VDO-related observational probes in centric relation differ in their longitudinal displacement within a common PCA representation?

The restriction to ONL, OC2.5, and OC3 was defined to focus on clinically comparable VDO-related constraints; however, this selection should be considered analytical and exploratory rather than confirmatory.

The relationship between occlusion, posture, and temporomandibular function remains debated \cite{manfredini2012occlusion,perinetti2006occlusion,michelotti2010occlusion}. Nevertheless, trigeminal afferents and stomatognathic inputs have been shown to interact with postural control systems in experimental contexts \cite{gangloff2002trigeminal,tardieu2009dental,deriu2003vestibulomasseteric}. This makes occlusal modulation a plausible constraint for exploratory analysis of postural and gait regulation.

\subsection{Gait acquisition}

Gait data were acquired using instrumented insoles, providing spatiotemporal and postural variables. Wearable insole systems have been used for gait assessment in neurological and ecological contexts, including Parkinson's disease \cite{delDin2016freeliving,mostovoy2023reliability,martin2024insoles}.

The dataset consisted of six CSV files corresponding to the three selected occlusal conditions across the two sessions:
\[
\text{M1-ONL}, \text{M1-OC2.5}, \text{M1-OC3}, \text{M2-ONL}, \text{M2-OC2.5}, \text{M2-OC3}.
\]

The retained variables included spatiotemporal parameters, asymmetry measures, center-of-pressure descriptors, and plantar capacitive indices. Each observation was represented as a multivariate vector:
\[
x_i = (x_{i1}, x_{i2}, \ldots, x_{in}).
\]

\subsection{Preprocessing and PCA projection}

Variables were harmonized across the six files. Categorical fields, absolute timestamps, empty technical columns, and non-informative fields were excluded from the numerical matrix. After these exclusions, 60 numerical biomechanical variables were retained and standardized prior to projection.

Principal Component Analysis (PCA) was used to construct an exploratory selected PCA representation:
\[
z_i = \Phi(x_i),
\]
where \(z_i\) corresponds to the low-dimensional representation of observation \(x_i\).

Unlike the revised Level~3 analysis, which used UMAP as an exploratory visualization of overlapping multivariate observations at M1, the present Level~4 analysis uses PCA to estimate longitudinal centroid displacement in a common linear coordinate system. PCA was selected for reproducibility and metric consistency across M1 and M2, not because it reveals an intrinsic latent organization. It remains a partial representation of the total variance \cite{jolliffe2016pca}.

The aim was not to infer a unique physiological selected PCA representation, but to obtain a common representational plane for comparing longitudinal displacement across conditions.

For each condition and session, the centroid of the projected observations was computed. The displacement between M1 and M2 centroids was used as an exploratory proxy for longitudinal representational stability:
\[
d_\lambda = \| \bar{z}_{M2,\lambda} - \bar{z}_{M1,\lambda} \|.
\]
Lower displacement was interpreted as an exploratory proxy for lower observed longitudinal reorganization, while larger displacement suggested stronger reorganization within the selected projection.

\subsection{Robustness analysis of centroid displacement}

Because centroid displacement constitutes the main operational proxy for observed longitudinal viability, an exploratory bootstrap analysis was performed to estimate the robustness of the observed displacement values. For each occlusal condition, observations from M1 and M2 were resampled with replacement within each session. Centroids were then recomputed in the selected PCA projection for each bootstrap iteration, and the corresponding M1--M2 centroid displacement was recalculated.

For each condition, this procedure generated an empirical distribution of longitudinal displacement values. The 2.5th and 97.5th percentiles of this distribution were reported as exploratory 95\% percentile intervals. The objective was not to perform confirmatory statistical inference, but to assess whether the observed ordering of centroid displacements remained stable under resampling of the available observations.

The bootstrap analysis was performed within the selected PCA projection and should therefore be interpreted as an internal robustness analysis of the present representation, not as a population-level confidence interval.

\subsection{Variance-normalized robustness analysis}

Because centroid displacement may be influenced by the dispersion of observations within each condition, an additional variance-normalized robustness analysis was performed.

Besides Euclidean centroid displacement, Mahalanobis distance was computed between the M1 and M2 centroids using the pooled covariance matrix of each occlusal condition. Unlike Euclidean distance, Mahalanobis distance accounts for the covariance structure of the latent representation and therefore normalizes for differences in within-condition dispersion.

This complementary analysis was not intended to replace the Level-4 viability proxy, but to evaluate whether the observed ordering remained stable after covariance normalization.

\section{Results}

\subsection{Dataset structure}

The six files included three occlusal conditions across two sessions. The number of observations was as follows: M1-ONL \(n=50\), M1-OC2.5 \(n=41\), M1-OC3 \(n=51\), M2-ONL \(n=60\), M2-OC2.5 \(n=60\), and M2-OC3 \(n=57\).

After exclusion of non-biomechanical fields, including absolute timestamps, categorical labels, empty technical columns, and non-informative technical fields, a total of 60 numerical biomechanical variables were retained for the exploratory PCA projection.

The first two principal components accounted for 17.6\% and 14.6\% of the variance, respectively, corresponding to 32.2\% of the total variance. These values indicate that the projection provides only a partial representation of the total variability, reinforcing the interpretation of the results as exploratory approximations rather than exhaustive representations of system dynamics.

\subsection{PCA representation and longitudinal centroid displacement}

\begin{figure}[H]
    \centering
    \includegraphics[width=0.98\textwidth]{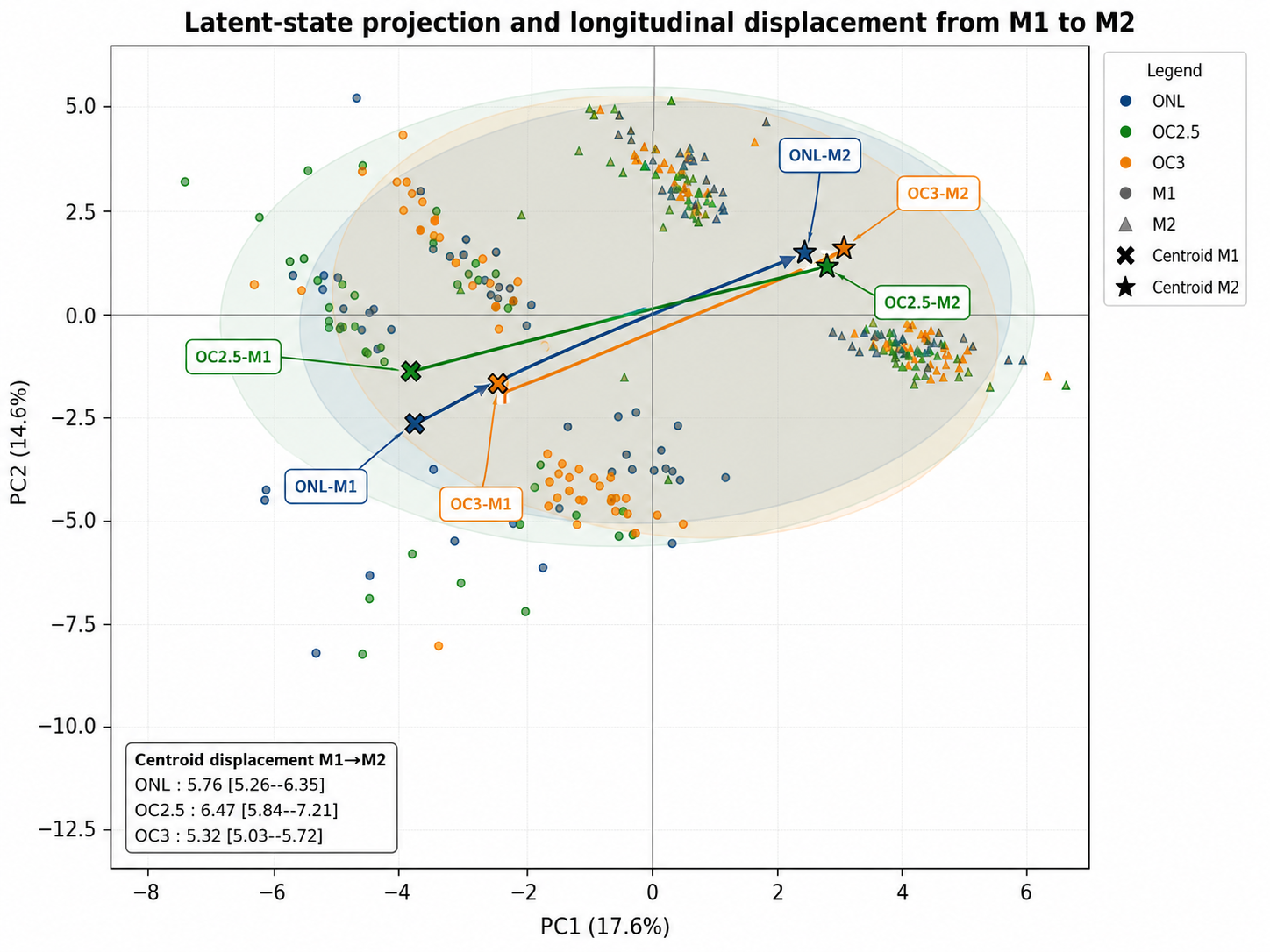}
    
    \caption{\textbf{Common PCA representation and longitudinal centroid displacement from M1 to M2.}
    PCA projection of gait observations obtained under the three selected observational probes (ONL, OC2.5, and OC3). The projection was computed from 60 standardized numerical biomechanical variables after exclusion of categorical fields, absolute timestamps, empty technical columns, and non-informative variables. PCA was fitted once to the combined M1--M2 analytical matrix, allowing all observations to be represented within the same common coordinate system. PC1 and PC2 accounted for 17.6\% and 14.6\% of the total variance, respectively. Points represent repeated gait observations from the single participant, colors identify the observational probes, and marker shapes distinguish sessions (M1 and M2). Condition centroids are shown for both sessions, with arrows indicating the observed longitudinal centroid displacement between M1 and M2 within the selected PCA representation. Bootstrap percentile intervals reported in the inset provide internal robustness estimates for the centroid displacements within this dataset. OC3 exhibited the lowest observed centroid displacement, OC2.5 the highest, and ONL an intermediate displacement. These centroid displacements describe representational changes within the selected PCA projection and should not be interpreted as direct physiological distances, intrinsic system stability, or causal effects of the occlusal probes.}
    
    \label{fig:pca}
    
    \end{figure}

The PCA projection revealed differentiated longitudinal displacements across occlusal conditions (Figure~\ref{fig:pca}). At M1, the centroids of ONL, OC2.5, and OC3 occupied a relatively close region of the selected PCA representation. Between M1 and M2, the conditions exhibited distinct patterns of longitudinal displacement in selected PCA representation.

The centroid displacement was:
\[
d_{\text{ONL}} = 5.76,
\qquad
d_{\text{OC2.5}} = 6.47,
\qquad
d_{\text{OC3}} = 5.32.
\]

As a complementary variance-normalized analysis, Mahalanobis distances were computed using the pooled covariance matrix of each condition. The results are shown in Table~\ref{tab:robustness}.

\begin{table}[H]
    \centering
    \small
    \caption{\textbf{Robustness analysis of the Level-4 viability proxy.}
    Euclidean centroid displacement preserves the ordering OC3 < ONL < OC2.5. After covariance normalization using Mahalanobis distance, the distances become closer and the ordering is not preserved. This indicates that within-condition covariance contributes to the observed latent displacement.}
    \label{tab:robustness}
    \begin{tabular}{lcccc}
    \hline
    Condition & Euclidean & Euclidean rank & Mahalanobis & Mahalanobis rank \\
    \hline
    OC3   & 5.32 & 1 & 1.893 & 3 \\
    ONL   & 5.76 & 2 & 1.814 & 2 \\
    OC2.5 & 6.47 & 3 & 1.809 & 1 \\
    \hline
    \end{tabular}
    \end{table}

Bootstrap resampling was used to estimate the robustness of these displacement values. The 95\% percentile intervals were:
\[
d_{\text{ONL}} = 5.76 \; [5.26--6.35],
\qquad
d_{\text{OC2.5}} = 6.47 \; [5.84--7.21],
\qquad
d_{\text{OC3}} = 5.32 \; [5.03--5.72].
\]

The relative ordering of centroid displacement was preserved in most bootstrap iterations, with OC3 showing the smallest displacement, OC2.5 the largest displacement, and ONL an intermediate displacement. The strict ordering
\[
d_{\text{OC3}} < d_{\text{ONL}} < d_{\text{OC2.5}}
\]
was observed in 85.7\% of bootstrap iterations. Pairwise comparisons showed:
\[
P(d_{\text{OC3}} < d_{\text{ONL}}) = 90.9\%,
\qquad
P(d_{\text{ONL}} < d_{\text{OC2.5}}) = 94.8\%,
\qquad
P(d_{\text{OC3}} < d_{\text{OC2.5}}) = 99.9\%.
\]

The bootstrap analysis indicates that the Euclidean Level-4 proxy is internally stable under resampling within the present dataset, without constituting confirmatory statistical evidence.

However, covariance normalization using Mahalanobis distance modified the relative ordering. This indicates that the current operational proxy is partly influenced by the covariance structure of the latent distributions. Rather than invalidating Level 4, this result identifies covariance-aware viability metrics as an important direction for future work.

OC3 exhibited the smallest centroid displacement in the selected PCA projection, showing comparatively lower longitudinal centroid displacement. OC2.5 showed the largest displacement, showing greater centroid displacement between M1 and M2 under the same projection. ONL occupied an intermediate position.

Because the bootstrap intervals partially overlapped, particularly between OC3 and ONL, this result should not be interpreted as a definitive clinical separation between conditions. Rather, it indicates an exploratory ordering tendency within the present single-case dataset and the selected projection.

Notably, OC2.5 and OC3 differ by only 0.5 degrees in vertical dimension of occlusion, yet exhibit different longitudinal displacements in the selected PCA projection. These differences indicate that configurations that appear close at M1 may follow different centroid displacements over time.

These interpretations remain based on a projected selected PCA representation and should be understood as approximations of the multivariate observations rather than direct measurements of trajectory stability.

The conceptual distinction introduced at Level 4 can be explored empirically by examining how configurations evolve between two time points under constraint. Although the selected occlusal conditions (ONL, OC2.5, OC3) exhibit comparable observable performance at baseline (M1), their longitudinal behavior differs when evaluated in selected PCA representation. The question is therefore not whether these configurations appear similar at a given time, but whether they evolve in a similar manner over time.

To address this question, the centroid displacement between M1 and M2 is used as an exploratory proxy for longitudinal behavior. This measure provides a coarse indication of how each configuration reorganizes under the combined effect of constraint and intervention.

Figure~\ref{fig:pca} visualizes these longitudinal displacements in the PCA-derived selected PCA representation, highlighting the divergence between configurations that appear initially close.

\subsection{A possible threshold between OC2.5 and OC3}

The most clinically relevant contrast concerns OC2.5 and OC3. These two conditions differ by only 0.5 degrees of VDO, yet they display distinct longitudinal behavior in the selected PCA projection. OC2.5 exhibits the largest centroid displacement, whereas OC3 exhibits the smallest displacement.

This suggests that the difference between 2.5 and 3 degrees should not be interpreted only as a simple linear increment in VDO. The contrast between OC2.5 and OC3 is compatible with a threshold-like interpretation, but does not establish the existence of a threshold. This interpretation remains exploratory and provides a rationale for Level 4 analysis: a small mechanical difference may correspond, under the present projection and in this single case, to a difference in longitudinal centroid displacement.

Overall, the results suggest that latent displacement may provide a discriminative signal between configurations that remain difficult to distinguish at the level of observable performance.

\paragraph{Result summary.}

Within the selected PCA projection, configurations that remained insufficiently resolved by static representations exhibited different condition-level centroid displacements between M1 and M2. OC3 showed a lower Euclidean displacement than ONL and OC2.5, whereas OC2.5 showed the largest displacement among the three selected probes. These findings describe representation-dependent longitudinal differences and should not be interpreted as direct evidence that the underlying physiological system evolved differently under each occlusal condition. This pattern supports Level~4 as an exploratory framework for describing longitudinal representational change.

\section{Discussion}

The present results support the relevance of a fourth analytical level centered on viability. While the revised Level~3 study showed that neither scalar aggregation nor a static exploratory embedding uniquely resolved the observational probes \cite{raynal2026level3}, the present study examines whether their condition-level representations exhibit different longitudinal displacements.
The variance-normalized robustness analysis further clarifies the scope of this interpretation. While the Euclidean centroid displacement preserved the ordering OC3 < ONL < OC2.5 under bootstrap resampling, Mahalanobis normalization did not preserve the same ordering. This suggests that within-condition dispersion contributes to the observed displacement. Therefore, the present Level-4 indicator should be interpreted as an exploratory proxy of observed longitudinal viability, not as a covariance-independent metric or validated biomarker. This limitation does not weaken the Level-4 framework itself, but clarifies that its operational implementation should evolve toward covariance-aware viability metrics.

The most clinically relevant finding concerns the comparative behavior of ONL, OC2.5, and OC3. In the selected PCA projection, the centroid displacement followed the ordering:
\[
d_{\text{OC3}} < d_{\text{ONL}} < d_{\text{OC2.5}}.
\]
Thus, OC3 showed the lowest longitudinal centroid displacement, ONL occupied an intermediate position, and OC2.5 showed the highest displacement.

Relative to ONL, OC3 exhibited a lower centroid displacement, whereas OC2.5 exhibited a higher centroid displacement. Quantitatively, OC3 showed an approximately 7.6\% lower displacement than ONL, while OC2.5 showed an approximately 12.3\% higher displacement than ONL. These values suggest that natural occlusion did not correspond to the lowest longitudinal reorganization in the selected projection, although the difference between OC3 and ONL should be interpreted as a tendency rather than a definitive clinical separation because the bootstrap intervals partially overlap.

The contrast between OC2.5 and OC3 is particularly informative. These two conditions differ by only 0.5 degrees of VDO, yet they occupy opposite positions in the displacement ranking. OC2.5 showed an approximately 21.6\% higher displacement than OC3. 

This observation suggests that a small mechanical difference in VDO may correspond, in this patient and under the selected projection, to a measurable difference in longitudinal representational behavior.

This pattern is compatible with a threshold-like interpretation, separating a pattern of greater reorganization from a pattern of relative stabilization. However, the present single-case design and the partial PCA representation do not establish the existence of a threshold or a dynamical regime transition. These values should not be interpreted as population-level effect sizes, but as descriptive within-subject contrasts in the selected PCA projection.

From a clinical perspective, instantaneous scalar or multivariate representations may leave several configurations unresolved. The present longitudinal analysis adds a complementary descriptor by comparing their M1--M2 centroid displacement. However, this descriptor is representation-dependent and does not establish that lower displacement corresponds to greater physiological stability, clinical benefit, or therapeutic suitability.

The sophrology-oriented sensorimotor intervention should be interpreted as a structured longitudinal context rather than as an isolated causal factor. Through body awareness, interoceptive attention, postural regulation, breathing control, and proprioceptive engagement, the intervention may have contributed to a modification of the participant's global sensorimotor organization between M1 and M2. Within the present Level 4 framework, its role is therefore not to explain the observed displacements directly, but to provide a structured transformation context in which the longitudinal behavior of each occlusal configuration could be observed.

Because the intervention was not controlled against a comparison condition, its specific contribution cannot be isolated from spontaneous evolution, occlusal constraint, or their possible interactions. The observed differences between ONL, OC2.5, and OC3 should therefore be interpreted as differential longitudinal behavior under a combined clinical context, not as direct effects of sophrology or occlusal adjustment alone.

These observations support a shift in clinical reasoning. The relevant question is not only which configuration produces acceptable performance at a given time, but which configuration is associated with lower observed longitudinal centroid displacement under transformation. This shift from state evaluation to trajectory evaluation is particularly relevant in adaptive neuromechanical systems, where compensation and reorganization are intrinsic properties.

Within this framework, occlusal adjustment should not be interpreted as a direct determinant of performance, but as a constraint that may participate in shaping the system's longitudinal organization. The objective is therefore not to identify a universally optimal occlusal value, but to examine how condition-level representations differ in their observed longitudinal displacement.

These interpretations must be considered in light of the exploratory single-case design and the use of PCA as an approximate latent representation; these limitations are detailed below.

Despite these limitations, the present work provides a structured framework for integrating longitudinal stability into clinical interpretation. By introducing observed longitudinal displacement as an exploratory criterion, it extends the analysis from what is observed at a given time to how a selected condition-level representation changes between sessions.

The companion Level-5 study extends this framework by introducing an internal predictive approximation of observed longitudinal viability, moving from retrospective observation toward predictive latent-space modeling \cite{raynal2026level5}.

\subsection{Clinical Interpretation: From Performance to Stability}

Clinical evaluation should not rely solely on instantaneous performance measures, but should also consider the longitudinal change of condition-level representations over time.

In the present case, occlusal conditions ONL, OC2.5, and OC3 occupied different positions in the longitudinal displacement ranking. OC3 showed the lowest centroid displacement, ONL occupied an intermediate position, and OC2.5 showed the highest displacement. Relative to ONL, OC3 showed lower longitudinal displacement, whereas OC2.5 showed higher longitudinal displacement. These differences suggest that close occlusal configurations may not support equivalent longitudinal representational behavior, even when instantaneous observable performance appears comparable.

This hierarchy supports the relevance of a Level 4 framework centered on longitudinal viability. The clinical question is therefore not only whether a configuration appears acceptable at a given time point, but whether it is associated with lower observed longitudinal centroid displacement over time.

From a clinical standpoint, this distinction is important. A configuration that appears acceptable at a given time may not support lower longitudinal reorganization. Conversely, a configuration associated with lower longitudinal displacement may be clinically relevant to examine further, even if immediate performance differences are minimal.

However, this interpretation does not establish a causal occlusal effect, a validated dynamical threshold, or a therapeutic optimum. The observed hierarchy should be understood as an exploratory within-subject finding in the selected PCA projection, not as a generalizable clinical rule.

Level 4 reframes the clinical question: not only which configuration performs best at a given time, but which configuration is associated with lower observed longitudinal centroid displacement over time.

\section{Limitations}

Several limitations must be emphasized. First, this study is based on a single participant and does not allow population-level generalization. Second, PCA provides a linear exploratory representation and does not capture the full complexity of the system. Third, centroid displacement is used here as a proxy for stability, but it does not constitute a validated clinical biomarker. Fourth, the sensorimotor intervention was not controlled against a comparison condition, and its effects cannot be isolated from spontaneous evolution or other contextual factors. Consequently, the specific contribution of the sensorimotor intervention cannot be isolated from the effects of occlusal constraint or natural longitudinal variability.

Although bootstrap resampling provided internal robustness estimates for centroid displacement, these intervals should not be interpreted as population-level confidence intervals. They quantify the stability of centroid displacement under resampling of the available observations within this single-case dataset. In addition, the reported distances depend on preprocessing choices, variable selection, standardization, and the number of retained PCA components.

Accordingly, the present work should be interpreted as a conceptual and exploratory extension, not as a causal demonstration. Its purpose is to develop a framework for interpreting viability and to illustrate this framework using real gait data.

Because the first two PCA components account for approximately one third of the total variance, the reported centroid displacements should be interpreted as displacements within the displayed projection, not as complete estimates of whole-system longitudinal change.

Finally, the Mahalanobis robustness analysis showed that covariance normalization partially modified the ranking obtained with Euclidean centroid displacement. Future work should therefore investigate covariance-aware or covariance-invariant viability metrics and compare alternative latent-space distances.

\section{Toward Level 5: Internal Predictive Approximation of Observed Longitudinal Viability}

The present paper remains at Level 4, where viability is assessed retrospectively through observed longitudinal centroid displacement. A natural extension of this framework would consist in predicting whether a given configuration is likely to exhibit stable longitudinal behavior under constraint.

Such an extension would require learning the transformation:
\[
\hat{z}_{t+1} = G_\theta(z_t,\lambda_t),
\]
and estimating whether the predicted state remains within the viability set \(\mathcal{V}\).

In this perspective, the objective is not merely to predict performance, but to anticipate whether a configuration is likely to maintain coherent longitudinal behavior over time. This connects clinical interpretation with representation learning and predictive modeling in selected PCA representation.

Recent approaches based on predictive representations and world models, including JEPA-like frameworks, provide a conceptual direction for modeling such dynamics directly in embedding space \cite{bengio2013representation,lecun2022path}.

However, this extension is not implemented in the present study. Level 5 should therefore be understood as a prospective direction, connecting viability assessment with predictive modeling.

Thus, Level 4 provides an observational framework for assessing observed longitudinal viability, whereas Level 5 introduces an internal predictive approximation of that observed viability through latent-space modeling.

Together, these two levels establish the transition from retrospective observation to predictive representation learning in adaptive biological systems.

\section{Conclusion}

This work introduces Level~4 as a longitudinal extension of a multi-level analytical framework. The revised Level~3 analysis showed that neither an aggregated scalar score nor a static exploratory UMAP embedding uniquely resolved the occlusal observational probes. Level~4 therefore addresses a different question: whether condition-level representations that remain ambiguous at one session exhibit different changes between M1 and M2.

In the selected PC1--PC2 representation, OC3 showed the lowest Euclidean centroid displacement, ONL occupied an intermediate position, and OC2.5 showed the highest displacement. This hierarchy was preserved under bootstrap resampling but was partly modified after covariance normalization. The result is therefore representation-dependent and influenced by within-condition dispersion.

These findings do not establish distinct physiological states, causal occlusal effects, a validated viability threshold, or a therapeutic optimum. Lower centroid displacement should not be equated directly with greater clinical or physiological stability.

The contribution of Level~4 is methodological: it introduces observed longitudinal displacement as a complementary descriptive property when static representations remain insufficient. What appears unresolved at one time point may nevertheless exhibit differentiated representational change over time.

The present work therefore provides a retrospective observational framework for longitudinal analysis while identifying covariance-aware or covariance-invariant metrics as necessary methodological developments. Its Level~5 extension examines whether the observed displacement can be internally approximated within the same single-subject dataset.

\section*{Author Contributions}

Jacques Raynal: conceptualization, study design, clinical occlusal assessment, definition and interpretation of occlusal constraints, computational analysis, AI-oriented interpretation framework, data analysis and writing-original draft.

Pierre Slangen: biomechanical interpretation, methodological review, and writing-review and editing.

Elsa Raynal: delivery of the sophrology sensorimotor protocol, clinical contextualization of the intervention within the dental and occlusal care context, contribution to the interpretation of sensorimotor regulation, and writing-review and editing.

Jacques Margerit: conceptual supervision, clinical and scientific interpretation, and writing-review and editing.

The present article proposes an exploratory observational implementation of Level~4 based on representation-dependent longitudinal centroid displacement. Its Level~5 extension examines whether this observed displacement can be internally approximated within the same single-subject dataset \cite{raynal2026level5}.

All authors reviewed and approved the final manuscript.

\section*{Data Availability}

The data analyzed in this study are not publicly distributed in the present preprint because they derive from a single clinical participant. Anonymized derived data or analysis outputs may be made available by the corresponding author upon reasonable request.

\end{document}